\documentclass[conference]{IEEEtran}
\IEEEoverridecommandlockouts

\usepackage[english]{babel}
\usepackage[T1]{fontenc}
\usepackage{amssymb, amsmath}
\usepackage{cite}
\usepackage{graphicx}
\usepackage{subfigure}

\newcommand{\norm}[2]{\| #1 \|_{#2}}
\newcommand{\cscgdist}[2]{\sim \mathcal{CN} \left( #1, #2 \right)}
\newcommand{\complexset}{\mathbb{C}}
\newcommand{\realset}{\mathbb{R}}
\newcommand{\integerset}{\mathbb{Z}}
\newcommand{\set}[1]{\left\{ #1 \right\}}

\def\BibTeX{{\rm B\kern-.05em{\sc i\kern-.025em b}\kern-.08em T\kern-.1667em\lower.7ex\hbox{E}\kern-.125emX}}

\begin{document}

\title{Reconfigurable Intelligent Surfaces to Enable Energy-Efficient IoT Networks}

\author{
\IEEEauthorblockN{Jo\~ao Henrique Inacio de Souza}
\IEEEauthorblockA{\textit{Electrical Eng. Dept.}\\
\textit{UEL}\\
Londrina, Brazil.\\
joaohis@outlook.com}

\and

\IEEEauthorblockN{Jos\'e Carlos Marinello Filho}
\IEEEauthorblockA{\textit{Electrical Eng. Dept.}\\
\textit{UTFPR}\\
Corn\'elio Proc\'opio, Brazil.\\
jcmarinello@utfpr.edu.br}

\and

\IEEEauthorblockN{Taufik Abr\~ao}
\IEEEauthorblockA{\textit{Electrical Eng. Dept.}\\
\textit{UEL}\\
Londrina, Brazil.\\
taufik@uel.br}

\and

\IEEEauthorblockN{Cristiano Panazio}
\IEEEauthorblockA{\textit{Telecommunications  Dept.}\\
\textit{EPUSP}\\
S\~ao Paulo, Brazil.\\
cpanazio@usp.br}

\thanks{This work was supported by the National Council for Scientific and Technological Development (CNPq) of Brazil under Grants 405301/2021-9, 141445/2020-3, and 310681/2019-7.}
}

\maketitle

\begin{abstract}
In this article, we study the uplink (UL) channel of a cellular network of Internet of Things (IoT) devices assisted by a reconfigurable intelligent surface (RIS) with a limited number of reflecting angle configurations.
Firstly, we derive an expression of the required transmit power for the machine-type devices (MTDs) to attain a target signal-to-noise ratio (SNR), considering a channel model that accounts for the RIS discretization into sub-wavelength reflecting elements. Such an expression demonstrates that the transmit power depends on the target SNR, the position of the MTD in the service area, and the RIS setup, which includes the number of reflecting elements and the available reflecting angle configurations.
Secondly, we develop an expression for the expected battery lifetime (EBL) of the MTDs, which explicitly depends on the MTD transmit power.
Numerical simulations on the energy efficiency (EE) evaluated via the EBL demonstrate the benefits of adopting RISs to enable energy-efficient IoT networks.
\end{abstract}


\section{Introduction}\label{sec:introduction}

RIS is a promising low-cost and low-power \textit{physical layer} (PHY) technology to enable sustainable wireless networks. This technology can produce controllable anomalous reflection of the incoming signal to create hot spots in areas of interest in the communication cell without any power amplifier (PA), consuming much less energy than common active PHY technologies \cite{liaskos2018}.
On the matter of EE, it is shown in \cite{huang2019} that RIS-assisted networks can achieve up to a threefold increase in EE when compared with networks assisted by multiple-antenna amplify-and-forward relays.
This benefit on EE is particularly favorable for IoT networks, which need to achieve low-power consumption and extended coverage to serve hundreds of battery-powered MTDs. In this sense, in \cite{cao2022, croisfelt2022}, efficient random access protocols are proposed for RIS-assisted networks, introducing frameworks to coordinate the optimization of the RIS reflecting coefficients with the device transmissions.
Moreover, in \cite{souza2022}, the EE of the random access channel in a RIS-assisted IoT network is evaluated, revealing that simultaneously optimizing the RIS setup and the transmission protocol is paramount to attain competitive EE and throughput combined with low-power consumption.

In this article, we study how RISs can improve the EBL of MTDs in IoT networks, focusing on the PHY aspects.
Firstly, we derive an expression for the required transmit power to attain a target SNR considering a channel model that accounts for the RIS discretization into sub-wavelength reflecting elements \cite{croisfelt2022}. Secondly, we develop an expression for the EBL of the MTDs, revealing that the numbers of reflecting elements and reflecting angle configurations at the RIS have significant impact on both system EE and network coverage.
Differently from the work \cite{souza2022}, in this article we focus on the EE problem exclusively at the devices-side, which needs more attention since the IoT devices are powered by limited-capacity batteries that are in general expensive and difficult to replace.
Numerical evaluations demonstrate that the RIS can significantly improve the EBL of the MTDs by reducing the transmit power. Nevertheless, since the EBL is simultaneously limited by the number of RIS elements and configurations, the RIS setup must be carefully designed to achieve energy-efficient networks that can fully exploit the potential of the RIS.

\section{System Model}\label{sec:system-model}

\begin{figure}[b!]
\vspace{-7mm}
\centering
\subfigure[Setup of the communication scheme]{
\includegraphics[width=.8\columnwidth]{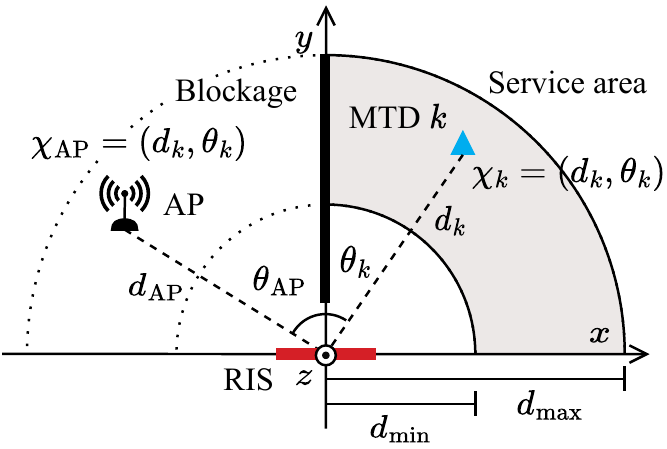}
\label{fig:communication-cell-setup}
}
\subfigure[TDMA frame]{
\includegraphics[width=\columnwidth]{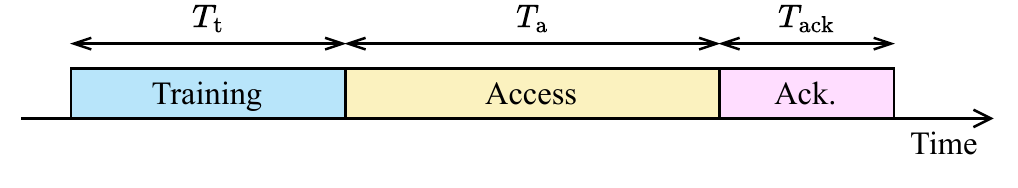}
\label{fig:tdma-frame}
}
\caption{a) AP, RIS, and service area of the communication cell where the MTDs are located; b) TDMA frame.}
\label{fig:tdma-frame-communication-cell-setup}
\end{figure}

In this section, we describe the system model. We consider a communication setup as depicted in Fig. \ref{fig:communication-cell-setup}, where the single-antenna MTDs are connected with the single-antenna AP via the reflected path produced by the RIS, since the direct radio links between the AP and the MTDs are highly attenuated due to a blockage.
In the communication scheme, the AP and MTDs are located at the $xy$-plane. Let $\chi_k = (d_k, \theta_k)$ denotes the position of the $k$-th MTD in the service area, where $d_{\min} \leq d_k \leq d_{\max}$ is the distance between the MTD and the origin, and $0 \leq \theta_k \leq \frac{\pi}{2}$ is the angle formed by the MTD with the RIS boresight. Similarly, let $\chi_{\textsc{ap}} = (d_{\textsc{ap}}, \theta_{\textsc{ap}})$ denotes the position of the AP.

The RIS is a thin surface located at the $xz$-plane with center at the origin. The surface is constituted by $N_x \in \integerset_+$ reflecting elements in the $x$-axis direction and $N_z \in \integerset_+$ in the $z$-axis direction, totaling $N = N_x N_z$ elements.
From the operational standpoint, we consider that the RIS introduces a phase shift $0 \leq \theta_r \leq \frac{\pi}{2}$ with marginal amplitude attenuation to the impinging signal. Importantly, the introduced phase shift belongs to a limited set of reflecting angle configurations, proposed to reduce the complexity of the RIS hardware and its control mechanism. Let $C \in \integerset_+$ denotes the number of available configurations, and $\delta \triangleq \frac{\pi}{2(C+1)}$ denotes the angular resolution of the RIS. The set of reflecting angle configurations is defined as:
\begin{equation}
\label{eq:set-reflecting-angle-configurations}
\Theta \triangleq \set{\theta_r \mid \theta_r = c \delta, \; c = 1,\dots,C}.
\end{equation}

In this article, we focus on the UL channel, assuming that the packet generation by each MTD is modeled as a Poisson process.
The UL transmissions are carried out through time-division multiple access (TDMA) frames with duration less than the channel coherence time, $T_c > 0$. The TDMA frame, depicted in Fig. \ref{fig:tdma-frame}, is comprised by a \textit{training} block of length $T_t > 0$, an \textit{access} block for data transmission with length $T_a > 0$, and an \textit{acknowledgement} block of length $T_{\text{ack}} > 0$.
In the training block, in coordination with the RIS, the AP broadcasts a signal to the MTDs to prepare for the data transmission. Then, the MTDs transmit their data signals in the access block. Lastly, the AP sends a message to recognize the successfully decoded data packets in the acknowledgement block.
To focus purely on the PHY aspects of the system, we adopt this generic description of the TDMA frame. For comprehensive information on designing transmission protocols for RIS-assisted networks for MTDs, see \cite{cao2022, croisfelt2022}.

During the access block, the received data signal of the $k$-th MTD at the AP considering single-device transmission is equal to:
\begin{equation}
\label{eq:received-signal}
\mathbf{y} \triangleq \sqrt{\rho_k} h_k \mathbf{v}_k + \mathbf{w},
\end{equation}
where $\rho_k > 0$ is the MTD transmit power, $h_k \in \complexset$ denotes the channel coefficient, $\mathbf{v}_k \in \complexset^{L}$ such that $\mathbb{E}[\norm{\mathbf{v}_k}{2}^2] = 1$ denotes the transmitted data signal with $L \in \integerset_+$ symbols, and $\mathbf{w} \in \complexset^L$ such that $\mathbf{w} \cscgdist{\mathbf{0}}{\sigma_w^2\mathbf{I}_L}$ denotes the additive white Gaussian noise vector, where $\sigma_w^2 > 0$ is the noise power.
From \eqref{eq:received-signal}, the instantaneous SNR for the $k$-th MTD is equal to:
\begin{equation}
\gamma_k \triangleq \frac{\rho_k |h_k|^2}{\sigma_w^2}.
\end{equation}
Based on the channel model of \cite{ozdogan2020, croisfelt2022} for communication schemes as depicted in Fig. \ref{fig:communication-cell-setup}, the SNR can be rewritten as:
\begin{equation}
\label{eq:snr}
\gamma_k = \frac{\rho_k \beta_k |A_k|^2}{\sigma_w^2},
\end{equation}
where $\beta_k > 0$ denotes the \textit{total path loss} and $A_k \in \complexset$ denotes the \textit{array factor} due to the RIS discretization into sub-wavelength reflecting elements \cite{croisfelt2022}, respectively defined as:
\begin{gather}
\label{eq:total-path-loss}
\beta_k = \frac{\beta_0}{d_{\textsc{ap}}^2 d_k^2} \cos^2 \theta_k,\\
\label{eq:array-factor}
A_k = N_z \sum_{n = 1}^{N_x} e^{j 2\pi (\sin \theta_k - \sin \theta_r) n},
\end{gather}
where $\beta_0 > 0$ denotes the path loss at the reference distance and $j = \sqrt{-1}$.
From \eqref{eq:array-factor}, one can notice that the RIS array factor directly depends on the reflecting angle $\theta_r$, which is limited by the set $\Theta$ defined in eq. \eqref{eq:set-reflecting-angle-configurations}.
Finally, substituting \eqref{eq:total-path-loss} and \eqref{eq:array-factor} in \eqref{eq:snr}, we obtain the following expression of the instantaneous SNR:
\begin{equation}
\label{eq:snr-complete}
\gamma_k = \frac{\rho_k \beta_0 N_z^2 \left| \sum_{n = 1}^{N_x} e^{j 2\pi (\sin \theta_k - \sin \theta_r) n} \right|^2 \cos^2 \theta_k}{\sigma_w^2 d_{\textsc{ap}}^2 d_k^2}.
\end{equation}
Therefore, one can note that, besides the MTD and AP positions, the RIS setup, specifically the number of reflecting elements, $N = N_x N_z$, and the reflecting angle configuration, $\theta_r \in \Theta$, has a significant impact on the instantaneous SNR.

\section{Device Energy Consumption Model}\label{sec:model-device-energy-consumption}

In this section, we develop a model for the MTDs energy consumption. We propose an expression for the EBL as EE metric specially focused on the devices-side.
Let $\bar{\gamma} \in \realset$ denotes the target SNR at the AP for an MTD to experience reliable communication. From \eqref{eq:snr-complete}, we can derive the transmit power of the $k$-th MTD to attain the average SNR $\bar{\gamma}$ as:
\begin{equation}
\label{eq:required-transmit-power}
\hspace{-2mm}\rho_k(\bar{\gamma}) = \bar{\gamma} \left( \frac{\beta_0 N_z^2 \left| \sum_{n = 1}^{N_x} e^{j 2\pi (\sin \theta_k - \sin \theta_r) n} \right|^2 \cos^2 \theta_k}{\sigma_w^2 d_{\textsc{ap}}^2 d_k^2} \right)^{-1}\hspace{-5mm}.\hspace{-3mm}
\end{equation}
From \eqref{eq:required-transmit-power}, one can notice that the transmit power depends on the position of the MTD, $\chi_k = (d_k, \theta_k)$, as well as on the number of elements at the RIS, $N = N_x N_z$, and the angle of reflection of the RIS, $\theta_r$. Deriving the radiofrequency (RF) transmit power is paramount to analyze the MTD EE as it constitutes most of the dissipated power during the TDMA 
frame.

To develop the model for the \textit{device energy consumption}, we consider the MTDs under two device modes: \textit{i)} \textit{transmission} mode, and \textit{ii)} \textit{receive} mode. From Fig. \ref{fig:tdma-frame}, one can perceive that the MTDs are in the receive mode during the training and acknowledgement blocks, while they are in the transmit mode during the access block. Therefore, the \textit{energy consumption} $E(\chi_k) > 0$ of the $k$-th MTD at the position $\chi_k$ is defined by:
\begin{equation}
E(\chi_k) \triangleq E_s + T_t P_{\text{Rx}} + T_a (P_c + \xi \rho_k) + T_{\text{ack}} P_{\text{Rx}},
\end{equation}
where $E_s > 0$ denotes the static energy consumption, $P_{\text{Rx}} > 0$ denotes the power dissipated in the receive mode, $P_{\text{Tx}} = P_c + \xi \rho_k$ denotes the power dissipated in the transmission mode, where $P_c > 0$ denotes the power dissipated by the device circuits, and $\xi > 1$ denotes the inverse of efficiency of the RF PA.

The EBL is a metric to evaluate the device EE given its battery capacity, energy consumption, and reporting period \cite{azari2021}. Let $T_r > 0$ denotes the expected length between consecutive \textit{reporting periods} of the MTD. The EBL in seconds of the $k$-th MTD at the position $\chi_k$ is equal to:
\begin{equation}
\label{eq:expected-battery-lifetime}
L(\chi_k) \triangleq \frac{E_0\, T_r}{E_s + T_a (P_c + \xi \rho_k) + (T_t + T_{\text{ack}}) P_{\text{Rx}}} ,
\end{equation}
where $E_0 > 0$ denotes the battery capacity of the MTD in Joule. From the transmission protocol standpoint, the EBL depends mainly on the length of the TDMA frame blocks and the transmit power. Especially, as suggested by \eqref{eq:required-transmit-power}, the transmit power is strongly linked to the RIS setup, particularly the number of RIS elements and reflecting angle configurations. Therefore, carefully designing the RIS setup is essential to improve the system EE in IoT networks, particularly at the devices-side.

\section{Numerical Results}\label{sec:numerical-results}

In this section, we analyze the EE of the RIS-assisted IoT network using the EBL as the main metric.
In the simulations, we calculate the required transmit power to attain the target SNR, defined in eq. \eqref{eq:required-transmit-power}, at each position $\chi_k$. Moreover, we assume that the transmit power cannot exceed the maximum transmit power, $\rho_{\max} > 0$, \textit{i.e.}, $\rho_k(\bar{\gamma}) \leq \rho_{\max}$. In this sense, when the MTD needs more power than $\rho_{\max}$ to attain the target SNR, such MTD is under outage.
All the parameter values deployed to generate the numerical results are listed in Table \ref{tab:numerical-results-parameter-values}.

\begin{table}[b]
\centering
\caption{Parameter values used to generate the numerical results.}
\label{tab:numerical-results-parameter-values}
\begin{tabular}{p{.28\columnwidth}|p{.6\columnwidth}}
\hline
\textbf{Parameter} & \textbf{Description}\\
\hline
$\chi_{\textsc{ap}} = (20 \text{m}, \pi/4)$ & AP position\\
$d_k \in [20, 100]$ m & Distance of the MTD from the origin\\
$\theta_k \in [0, \pi/2]$ & MTD-RIS angle\\
\hline
${N \in [4,100]}$ & \# RIS elements ($N_x = N_z$)\\
$C \in \set{2,4,8,16}$ & \# RIS reflecting angle configurations\\
\hline
$T_c = 50$ ms & Channel coherence time\\
$T_a = 0.85 T_c$ & Training block length\\
$T_t = 0.10 T_c$ & Access block length\\
$T_{\text{ack}} = 0.05 T_c$ & Acknowledgement block length\\
$\beta_0 = - 52$ dB & Path loss at the reference distance\\
$\sigma_w^2 = -94$ dBm & Noise power\\
\hline
$T_r = 300$ s & Length between consecutive reporting periods\\
$E_0 = 2.5$ kJ & Battery capacity\\
$E_s = 10 \mu$J & Static energy consumption\\
$P_c = 1 $ mW & Power dissipated by the MTD circuits\\
$P_{\text{Rx}} = 100$ mW & Power dissipated in the receive mode\\
$\xi = 1.33$ & Inverse of efficiency of the RF PA\\
$\rho_{\max} = 24$ dBm & Maximum transmit power\\
$\bar{\gamma} = 10$ dB & Target SNR\\
\hline
\end{tabular}
\end{table}

\begin{figure}[t]
\centering
\includegraphics[width=\columnwidth]{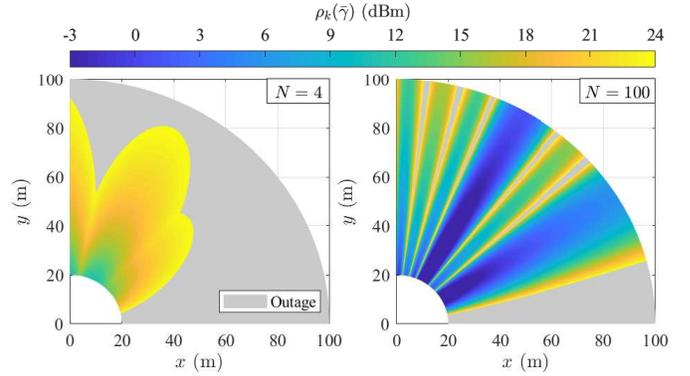}
\vspace{-7mm}
\caption{$\rho_k(\bar{\gamma})$ across the service area. $C = 2$, $\bar{\gamma} = 10$ dB, and $\rho_{\max} = 24$ dBm.}
\label{fig:power}
\vspace{-3mm}
\end{figure}

Fig. \ref{fig:power} depicts $\rho_k(\bar{\gamma})$ required to attain the target SNR $\bar{\gamma} = 10$ dB across the service area of the communication cell. We assume that the RIS has two different reflecting angle configurations, \textit{i.e.}, $C = 2$. Additionally, at each position, $\rho_k(\bar{\gamma})$ is calculated considering the best RIS configuration, that is, the configuration that results in the smallest transmit power.
Analyzing Fig. \ref{fig:power}, one can see that the minimum $\rho_k(\bar{\gamma})$ observed across the service area decreases with $N$ due to the enhanced array factor of the RIS, defined in eq. \eqref{eq:array-factor}. At the same time, the outage region area reduces as $N$ increases, improving the network coverage.
For $N = 100$, there are two main beams in the service area (dark blue color), where the required $\rho_k(\bar{\gamma})$ is less than the values in the other positions. Specifically, these beams are pointed towards the reflecting angles configured at the RIS, $\Theta$, which in the analyzed scenario with $C = 2$ are equal to $30^\circ$ and $60^\circ$. Hence, these beams represent the regions where the RIS can efficiently reflect much of the signal power.

\begin{figure}[t]
\centering
\subfigure[$C = 2$]{
\includegraphics[width=\columnwidth]{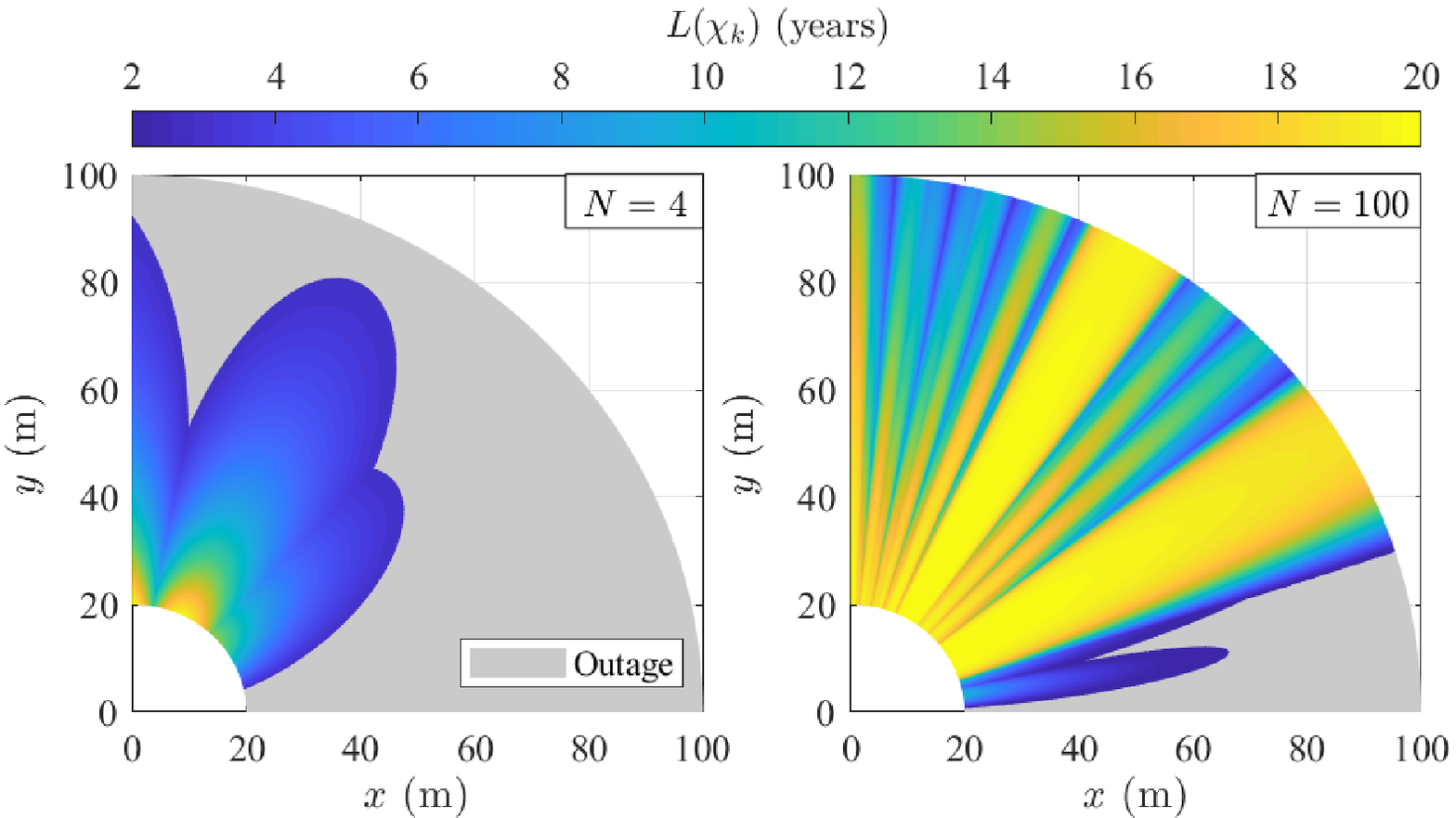}
\label{fig:ebl-C-2}
}
\subfigure[$C = 8$]{
\includegraphics[width=\columnwidth]{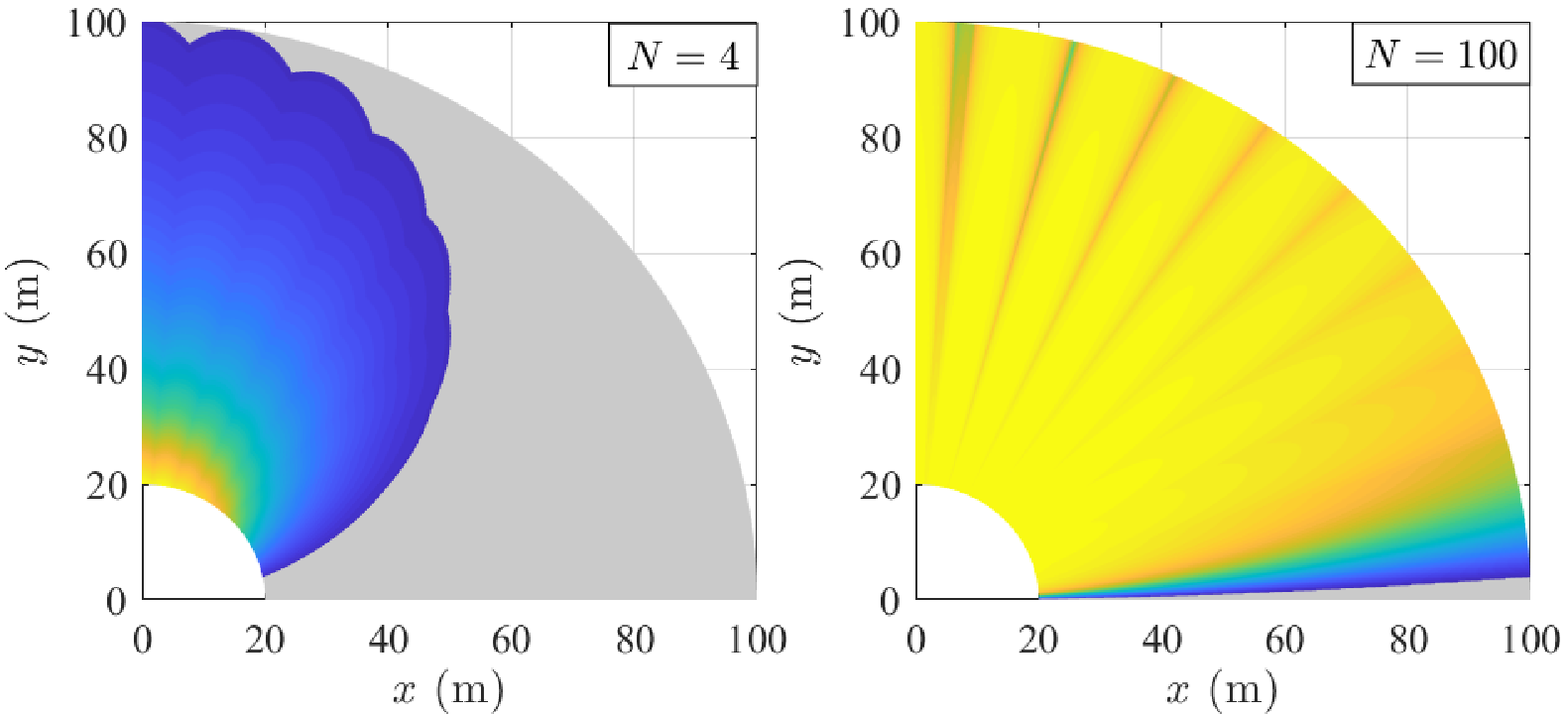}
\label{fig:ebl-C-8}
}
\vspace{-5mm}
\caption{$L(\chi_k)$ across the service area. $C \in \set{2,8}$, $T_r = 300$ s, $\bar{\gamma} = 10$ dB, and $\rho_{\max} = 24$ dBm.}
\vspace{-5mm}
\end{figure}

Based on the results of Fig. \ref{fig:power}, Fig. \ref{fig:ebl-C-2} depicts the EBL $L(\chi_k)$ in years across the service area for $C = 2$. From \eqref{eq:expected-battery-lifetime}, it is straightforward to see that $L(\chi_k)$ must follow the inverse behavior of $\rho_k(\bar{\gamma})$. Accordingly, in Fig. \ref{fig:ebl-C-2}, one can notice that the increase on the RIS array factor due to the higher $N$ improves the maximum $L(\chi_k)$, reducing the outage region area, and simultaneously improving the devices EE and the network coverage.

In order to analyze the effect of increasing the number of RIS reflecting angle configurations, Fig. \ref{fig:ebl-C-8} depicts $L(\chi_k)$ across the service area for $C = 8$.
Comparing this result with the same $N$ to Fig. \ref{fig:ebl-C-2}, one can notice that increasing $C$ improves the EBL $L(\chi_k)$ across the service area and reduces the outage region size. Such improvements arise since the RIS with an increased number of reflecting angle configurations can beamform the signal towards a greater number of directions.
Despite the improvement on coverage, one can notice that specifically for $N = 100$ there is a permanent outage region into the service area near to $\theta_k = \frac{\pi}{2}$. Such region is due to the setup of the communication cell, depicted in Fig. \ref{fig:communication-cell-setup}, where the MTDs with $\theta_k\approx\frac{\pi}{2}$ are located in approximately the same plane as the RIS, an unfavorable region for signal reflection as revealed by the cosine term in \eqref{eq:total-path-loss}.

\begin{figure}[b]
\vspace{-3mm}
\centering
\subfigure[Average $\rho_k(\bar{\gamma})$ and $L(\chi_k)$]{
\includegraphics[width=.492\columnwidth]{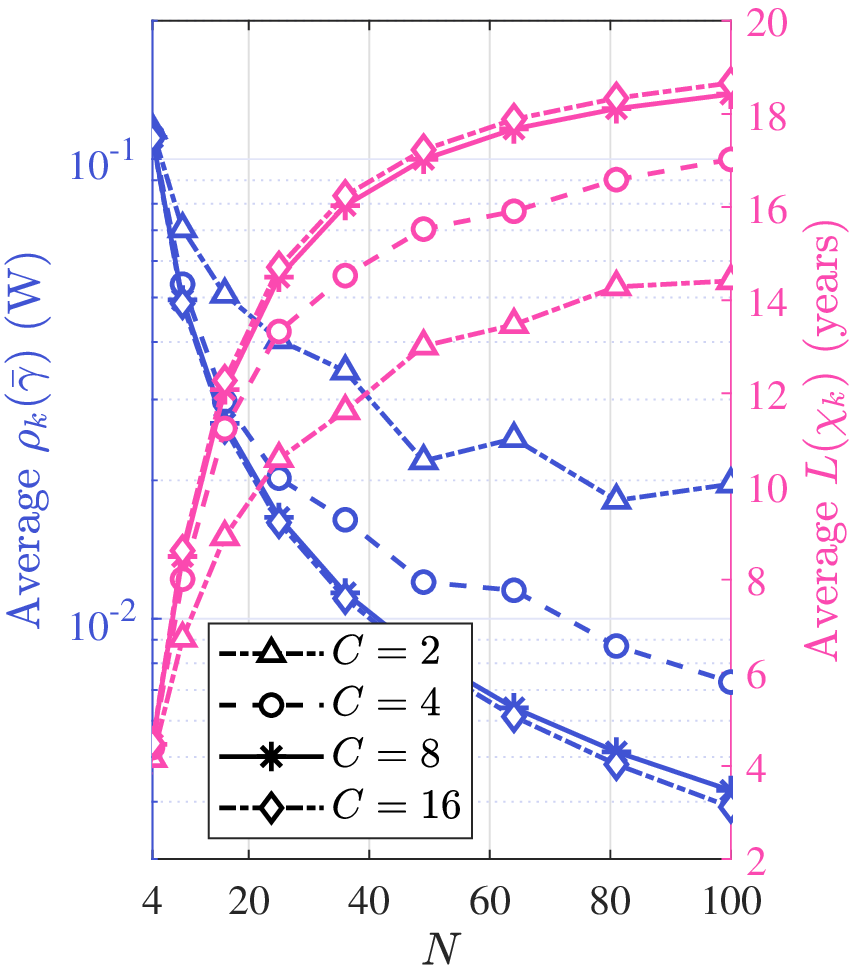}
\label{fig:average-power-and-ebl}
}
\subfigure[Outage region area]{
\includegraphics[width=.428\columnwidth]{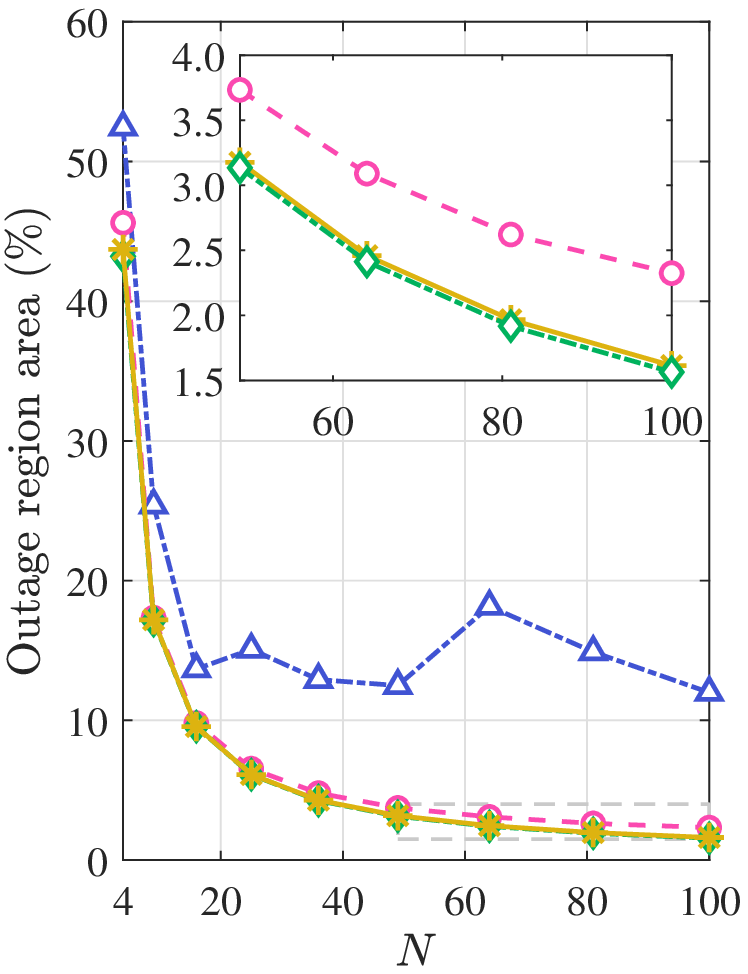}
\label{fig:outage-region-area}
}
\caption{a) Average $\rho_k(\bar{\gamma})$ and average $L(\chi_k)$ as a function of $N$; b) Outage region area as a function of $N$. $C \in \set{2,4,8,16}$, $T_r = 300$ s, $\bar{\gamma} = 10$ dB, and $\rho_{\max} = 24$ dBm.}
\end{figure}

Fig. \ref{fig:average-power-and-ebl} depicts the average $\rho_k(\bar{\gamma})$ and $L(\chi_k)$ over the service area as a function of $N$. Following the behavior observed in the previous results, the average $\rho_k(\bar{\gamma})$ decreases with $N$, while the average $L(\chi_k)$ increases with it.
Specifically for $C \in \set{2,4}$ and $N > 36$, one can notice that the average $\rho_k(\bar{\gamma})$ has a clear oscillating behavior. This is due to the low number of main beams produced by the small number of RIS reflecting angle configurations $C$ and the fact that increasing $N$ may increase the signal power at the main beams at the expense of decreasing it at the side beams.
Moreover, for a fixed $C$, one can notice that the average $\rho_k(\bar{\gamma})$ and, consequently, the average $L(\chi_k)$ are limited by $N$, demonstrating that the RIS setup, \textit{i.e.}, $N$ and $C$, must be carefully designed to avoid a setup that cannot be fully exploited by the network.

Lastly, Fig. \ref{fig:outage-region-area} depicts the percentage of the outage region area as a function of $N$. This result reinforces that the outage region decreases asymptotically both with $N$ and $C$, indicating that the RIS setup is directly associated with the network coverage. Indeed, such a result demonstrates that, even with large values of $N$ and $C$, there is a permanent outage region for $\theta_k \approx \frac{\pi}{2}$ due to the communication cell setup. Therefore, if the aim is to enhance the network coverage in the simulated setup, a RIS with $(N,C) = (81,8)$ is sufficient to get an outage region with area as small as $2\%$.

\section{Conclusions}\label{sec:conclusions}

In this article, we study the design of energy-efficient IoT networks assisted by a RIS. Firstly, we derive an expression for the required transmit power to attain a target SNR considering the RIS discretization into sub-wavelength reflecting elements. Secondly, we develop an expression for the EBL of the MTDs.
The numerical results demonstrate that the RIS can significantly improve the EBL of the MTDs by reducing the transmit power. Additionally, the RIS can enhance the network coverage as long as it has the right numbers of reflecting elements and reflecting angle configurations.
Nevertheless, since the EBL is simultaneously limited by the number of RIS elements and configurations, the RIS setup must be carefully designed to achieve energy-efficient networks that can fully exploit the potential of the RIS.

\bibliographystyle{IEEEtran}
\bibliography{references}

\end{document}